\documentclass[12]{article}
\usepackage{dcolumn}
\usepackage{color}
\usepackage{latexsym}
\usepackage{bm}
\begin{document}
\date{}
%%%%%%%%%%%%%%%%%%%%%%%%%%%%%%%%%%%%%%%%%%%%%%%%%%%%%%%%%%%%%%%%%%%%%%%%%%
%\title{\textbf{Covariant Anomalies and Hawking Radiation}}
%\author{{Rabin Banerjee}\thanks{E-mail: rabin@bose.res.in}, \ {Shailesh Kulkarni}\thanks{E-mail: shailesh@bose.res.in}\\
%\\\textit{S.~N.~Bose National Centre for Basic Sciences,}
%\\\textit{JD Block, Sector III, Salt Lake, Kolkata-700098, India}}
%%%%%%%%%%%%%%%%%%%%%%%%%%%%%%%%%%%%%%%%%%%%%%%%%%%%%%%%%%%%%%%%%%%%%%%%%%%%%%
\title{Hawking Radiation, Effective Actions and Covariant Boundary Conditions}
\author{{Rabin Banerjee}$^{1,2}$\thanks{E-mail: rabin@bose.res.in}, \  {{Shailesh Kulkarni}$^2$\thanks{E-mail: shailesh@bose.res.in}}\\
$^1$\textit{Institute of Quantum Science, College of Science and}\\
\textit{Technology, Nihon University, Tokyo 101-8308, Japan.}\\
$^2$\textit{S.N. Bose National Centre for Basic Sciences,}\\
\textit{JD Block, Sector III, Salt Lake, Kolkata-700098, India}}

%%%%%%%%%%%%%%%%%%%%%%%%%%%%%%%%%%%%%%%%%%%%%%%%%%%%%%%%%%%%%%%%%%%%%%%%%%%%%
%\begin{abstract}
\maketitle
\begin{quotation}
\noindent \normalsize
From an appropriate expression for the effective action, the Hawking radiation from charged black
holes is derived, using only covariant boundary conditions at the event horizon. The connection
of our approach with the Unruh vacuum and the recent analysis \cite{Robwilczek,Isowilczek,shailesh}
of Hawking radiation using anomalies is established.
\end{quotation}
%\end{abstract}
%\maketitle
{\it Introduction:}

Hawking radiation arises upon the quantisation of matter in a background spacetime with an event horizon.
It therefore plays an important role in black hole physics. Apart from Hawking's \cite{Hawking} original derivation,
there are  other approaches \cite{gibbons,parikh}, although none is completely clinching or conclusive.
This has led researchers to consider alternative derivations providing new insights into the problem.
Here we discuss another approach that is based solely on the structure
of the effective action and boundary conditions at the event horizon. We therefore guarantee the universality of Hawking radiation which ought to be
determined by properties at the event horizon only- a feature that is usually lacking in approaches based
on effective actions \cite{Fulling,wipf}. To put our work in a proper
perspective, however, it is desirable to elaborate on the recent approaches \cite{Robwilczek,Isowilczek,shailesh} to the 
Hawking effect which rely
on the cancellation of gauge and gravitational anomalies.

An anomaly, it might be recalled, is a breakdown of some classical symmetry due to the process of quantisation.
For example, a gauge anomaly is an anomaly in  gauge symmetry, taking the form of nonconservation of the gauge current.
Similarly, a gravitational anomaly occurs from a breaking of general covariance, taking the form of nonconservation
of the energy momentum tensor (For reviews see, \cite{bertlmann,Fujikawa}). The simplest manifestation of these 
(gauge and gravitational) anomalies, which is also relevant for the present discussion, occurs for $1+1$ dimensional
chiral fields.

Recently, Robinson and Wilczek \cite{Robwilczek}, followed by Iso, Umetsu and Wilczek \cite{Isowilczek}, gave a new derivation of the Hawking effect.
They found that, by the process of 
dimensional reduction,  effective field theories become two dimensional and chiral near the event horiazon of a black hole.
This leads to the occurrence of gauge and gravitational anomalies. The Hawking flux is necessary to cancel these
anomalies.

An essential aspect of  \cite{Robwilczek,Isowilczek} is that a two dimensional chiral theory admits two types of anomalous currents (and/or
energy momentum tensors)- the consistent and the covariant \cite{bertlmann,Fujikawa,Witten,Bardeen,rabin2,rabin1}-
which are actually related by local counterterms.  The covariant divergence of these currents and energy 
momentum tensors yields either the consistent or covariant form of the anomaly. Then the Hawking flux was derived
in \cite{Robwilczek,Isowilczek} by a cancellation of the consistent anomaly but the boundary condition necessary to fix the parameters was
obtained from a vanishing of the covariant current at the horizon \cite{Isowilczek}. It was also observed \cite{Morita} that 
an incorrect result for the charge flux would be obtained if, instead, the vanishing of the consistent current at
the horizon was taken as the boundary condition.

 The approach of \cite{Robwilczek,Isowilczek} was very recently generalised by us \cite{shailesh}. It was shown that the complete analysis was feasible
in terms of covariant expressions only. The flux from a charged black hole was correctly determined by a cancellation
of the covariant anomaly with the boundary condition being the vanishing of the covariant current 
(and energy momentum tensor) at the horizon.
Apart from being conceptually clean and more natural (all expressions being covariant), it simplified the original
analysis \cite{Robwilczek,Isowilczek} considerably. This was true not just for charged black holes, but for other black holes as well \cite{peng1}.

From the analysis of \cite{Isowilczek,shailesh} it appears therefore that covariant boundary conditions at the horizon play a fundamental
role. We adopt the arguments of \cite{Robwilczek,Isowilczek} which imply that effective
field theories are chiral and two dimensional near the horizon. Then, exploiting known structures of the two dimensional 
effective actions, the relevant expressions for the currents and the energy momentum tensors are derived by only imposing covariant
boundary conditions at the horizon. The Hawking flux from charged black holes is correctly reproduced in this manner. 
 Finally,
we establish the connection of our approach with calculations based on the Unruh vacuum \cite{unruh, Isoumtwilczek}.

{\it General Setting and Effective Actions :}\\

We are interested in discussing the Hawking effect from a charged black defined by the Reissner-Nordstrom metric
given by,
\begin{equation}
ds^2 = f(r) dt^2 - \frac{1}{f(r)} dr^2 -r^2d\Omega^2_{(d-2)}
\label{1.1}
\end{equation}
where $d\Omega^2_{(d-2)}$ is the line element on the $(d-2)$ sphere. The function
$f(r)$ admits an event horizon at $r_+$ so that $f(r_+)=0$. The gauge potential is defined 
by $A=-\frac{Q}{r} dt$.

As already mentioned by using a dimensional reduction technique,
the effective field theory  near the 
 horizon becomes a two dimensional chiral theory. The metric of this two dimensional
theory is identical to the $r- t$ component of the full metric (\ref{1.1}).
Hence the determinant of the metric simplifies to unity $(\sqrt{(-g)} = 1)$ and many expressions mimic
their flat space counterparts. The theory away from the horizon is not chiral and hence is anomaly free.

We now summarise, step by step, our methodology.
For a two dimensional theory the expressions for the effective actions, whether anomalous
(chiral) or normal, are known \cite{Isoumtwilczek, Leut}. Both these are required in our analysis.
For deriving the Hawking flux, only the form of the anomalous (chiral) effective action, which describes the theory
near the horizon, is required.
The currents and energy momentum tensors are  computed by
taking appropriate functional derivatives of this effective action. Next, the parameters appearing
in these solutions are fixed by imposing the vanishing of covariant currents (energy momentum
tensors) at the horizon. Once these are fixed, the Hawking fluxes
 are  obtained from the asymptotic $(r\rightarrow {\infty})$ limits of the currents and energy momentum
tensors. To show the connection with the Unruh vacuum the form of the usual effective action, which describes the
theory away from the horizon, is necessary. The currents and energy momentum tensors, obtained from this effective action,
are solved by using the knowledge of the corresponding chiral expressions. The results reproduce the expectation values
of the currents and energy  momentum tensors for the Unruh vacuum.

First, we consider the effective theory
 away from  the horizon. This is defined by the standard effective action $\Gamma$ of a conformal field
 with a central charge $c=1$ in this blackhole background \cite{Isoumtwilczek}.
 $\Gamma$  consists of two parts; the gravitational (Polyakov) part  and
 the gauge part. Adding the two contributions we obtain,
\begin{eqnarray}
\Gamma =  \frac{1}{96\pi}\int d^2x d^2y \ \sqrt{-g} R(x)\frac{1}{\Delta_{g}}(x,y)\sqrt{-g}R(y) + \nonumber\\
 \frac{e^2}{2\pi} \int d^2x d^2y  \ \epsilon^{\mu\nu}\partial_{\mu}A_{\nu}(x)\frac{1}{\Delta_{g}}(x,y) \epsilon^{\rho\sigma}\partial_{\rho}A_{\sigma}(y).
\label{1}
\end{eqnarray}
Here $R$ is the two-dimensional Ricci scalar given by $R=f''$, and 
and $\Delta_{g} = \nabla^{\mu}\nabla_{\mu}$ is the laplacian in this 
background.

    The energy-momentum tensor $T_{\mu\nu(o)}$ in the region outside 
the horizon is defined as,
\begin{equation}
T_{\mu\nu(o)} = \frac{2}{\sqrt{-g}} \frac{\delta\Gamma}{\delta g^{\mu\nu}}.
\label{2}
\end{equation}
The explicit form for $T_{\mu\nu(o)}$ is thus given by
\begin{eqnarray}
T_{\mu\nu(o)} &=& \frac{1}{48\pi} \left(2g_{\mu\nu}R - 2\nabla_{\mu}\nabla_{\nu}G + \nabla_{\mu}G \nabla_{\nu}G -\frac{1}{2}g_{\mu\nu}\nabla^{\rho}\nabla_{\rho}G\right)\nonumber\\
&& + \frac{e^2}{\pi} \left(\nabla_{\mu}B\nabla_{\nu}B- \frac{1}{2}g_{\mu\nu}\nabla^{\rho}B\nabla_{\rho}B\right) \label{3}
\end{eqnarray}
Similarly, the form for the gauge current can be obtained,
\begin{equation}
J^{\mu}_{(o)} =\frac{\delta\Gamma}{\delta A_{\mu}} = 
 \frac{e^2}{\pi} \epsilon^{\mu\nu}\partial_{\nu}B. \label{4}
\end{equation}
Here
%\begin{equation}
\begin{eqnarray}
G(x) &=& \int d^2y  \ \Delta^{-1}_{g}(x,y)\sqrt{-g}R(y),\label{5}\\
%\begin{equation}
B(x) &=& \int d^2y  \  \Delta^{-1}_{g}(x,y)\epsilon^{\mu\nu}\partial_{\mu}A_{\nu}(y).\label{5a}
\end{eqnarray}
%\end{equation}
From now on we would omit the $\sqrt{-g} =1$ factor from all the expressions. Hence we work with
the antisymmetric numerical tensor $\epsilon^{\mu\nu}$ defined by $\epsilon^{tr}=1$.
$B(x)$ and $G(x)$ satisfy,
\begin{equation}
\nabla^{\mu}\nabla_{\mu}B = - \partial_{r}A_{t} \ ; \  \nabla^{\mu}\nabla_{\mu}G = R = f'', \label{6}
\end{equation}
respectively. The solutions for $B$ and $G$ are now given by
\begin{equation}
B = B_o(r) - at + b \ ; \  \partial_{r}B_o = \frac{1}{f}(A_{t}+c),\label{7}
\end{equation}
\begin{equation}
G = G_o(r) - 4 pt + q \ ; \  \partial_{r}G_o = - \frac{1}{f}(f'+z),\label{8}
\end{equation}
where $a,b, c, p, q $ and $z$ are constants. Also note that $B_o$ and $G_o$ are functions of
$r$ only.

The current (\ref{4}) and the energy momentum tensor (\ref{3}) satisfy the normal Ward identities,
\begin{equation}
\nabla_\mu J^\mu_{(o)} = 0 \,\,\,; \,\,\, \nabla_\mu T^\mu_{\nu(o)}=F_{\mu\nu}J^\mu_{(o)}
\label{1.2}
\end{equation}
Note that in the presence of an external gauge field the energy momentum tensor is not conserved; rather
the Lorentz force term is obtained.

      In the region near the horizon we have gravitational as well
 as gauge anomaly so that the effective theory is described by an anomalous
 (chiral) effective action which is given by \cite{Leut}, 
\begin{equation}
\Gamma_{(H)}= -\frac{1}{3} z(\omega) + z(A)
\label{effaction}
\end{equation}
where $A_{\mu}$ and $\omega_{\mu}$ are the gauge field and the spin connection, respectively, and,
\begin{equation}
z(v) = \frac{1}{4\pi}\int d^2x d^2y \epsilon^{\mu\nu}\partial_\mu v_\nu(x) \nabla^{-1}(x, y)
\partial_\rho[(\epsilon^{\rho\sigma} + g^{\rho\sigma})v_\sigma(y)]
\label{effaction1}
\end{equation}

From a variation of this effective action the energy momentum tensor and the gauge current are computed.
To get their covariant forms in which we are interested, however, appropriate local polynomials have to be
added. This is possible since energy momentum tensors and currents are only defined modulo local polynomials.
The final results for the covariant energy momentum tensor and the covariant 
current are given by \cite{Leut}, 
\begin{equation}
\delta \Gamma_{H} = \int d^2x  \left( \frac{1}{2}\delta g_{\mu\nu} T^{\mu\nu} + \delta A_{\mu}J^{\mu}\right) 
+  l \label{chiralaction} 
\end{equation}
where the local polynomial is given by,
\begin{equation}
 l = \frac{1}{4\pi}\int d^{2}x \ \epsilon^{\mu\nu}(A_{\mu}\delta A_{\nu} - 
 \frac{1}{3}w_{\mu}\delta w_{\nu} - \frac{1}{24}R e_{\mu}^{a}\delta e_{\nu}^{a})
\end{equation}
The covariant energy momentum tensor $T^{\mu\nu}$ and the covariant gauge current $J^{\mu}$ are obtained from
the above relations as,
\begin{eqnarray}
T^{\mu}_{\nu} = \frac{e^2}{4\pi}\left(D^{\mu}B D_{\nu}B\right) \nonumber\\
 +\frac{1}{4\pi}\left(\frac{1}{48}D^{\mu}G D_{\nu}G 
-\frac{1}{24} D^{\mu} D_{\nu}G + \frac{1}{24}\delta^{\mu}_{\nu}R\right)
\label{9} 
\end{eqnarray}
 \begin{equation}
J^{\mu} = -\frac{e^2}{2\pi}D^{\mu}B.\label{10}
\end{equation}
Note the presence of the chiral covariant derivative,
\begin{equation}
D_{\mu} = \nabla_{\mu} - \epsilon_{\mu\nu}\nabla^{\nu} = -\epsilon_{\mu\nu}D^{\nu}, \label{11}
\end{equation} 
instead of the usual one that occurred previously in (\ref{3}), (\ref{4}). The definitions of $B$ and $G$ are provided in (\ref{5}), (\ref{5a}). 

By taking the covariant divergence of (\ref{9}) and 
(\ref{10}) we get the anomalous Ward identities,
\begin{equation}
\ \nabla_{\mu}J^{\mu} = -\frac{e^2}{2\pi} \epsilon^{\rho\sigma}\partial_{\rho}A_{\sigma} = \frac{e^2}{2\pi}\partial_{r}A_{t}\label{12}
\end{equation}
\begin{equation}
 \nabla_{\mu}T^{\mu}_\nu = F_{\mu\nu}J^{\mu} + \frac{1}{96\pi} \epsilon_{\nu\mu}\partial^{\mu}R.\label{13}
\end{equation}

The anomalous terms are the covariant gauge anomaly and the covariant gravitational anomaly, respectively. These
Ward identities were also obtained from different considerations in \cite{shailesh}. 
\newpage
{\it Charge and Energy Flux:}\\

In this subsection we calculate the charge and energy flux by using, respectively, the 
 expressions for the covariant current (\ref{10}) and the covariant energy momentum tensor (\ref{9}).
We will see that the results are the
 same as that  obtained by the anomaly cancellation (consistent or covariant) method \cite{Isowilczek,shailesh}. 
 
      First, we derive the charge flux. Using (\ref{7}) and 
 (\ref{11}) we have from (\ref{10}),
\begin{equation}
J^{r} = \frac{e^2}{2\pi}\left(A_{t}(r) + c + a\right) 
\label{14}
\end{equation}
 We now impose the boundary condition that the covariant current $J^{r}$ vanishes at  the  horizon,
implying $J^{r}(r_+)=0$.
This leads to a relation,
\begin{equation}
 c + a = - A_{t}(r_{+})
\label{coeff}
\end{equation}
Hence the expression for $J^{r}$  takes the form,
\begin{eqnarray}
J^{r} =\frac{e^2}{2\pi}\left(A_{t}(r) - A_{t}(r_{+})\right)
 \label{15}
\end{eqnarray}

Now the charge flux is given by the asymptotic $(r \rightarrow \infty)$ limit of the anomaly free current \cite{Robwilczek, 
Isowilczek, shailesh}. As observed from (\ref{12}) the anomaly vanishes in this limit and hence we directly compute the
flux from (\ref{15}) by taking the $(r \rightarrow \infty)$ limit. This yields,
\begin{equation}
J^{r}( r \rightarrow \infty) = - \frac{e^2}{2\pi}\left(A_{t}(r_+) \right) 
\label{chargeflux}
\end{equation}
This is the desired Hawking flux and agrees with previous findings \cite{Robwilczek, 
Isowilczek, shailesh}. 

We next consider the energy momentum flux by adopting the same technique. After using the solutions
 for $B(x)$ and $G(x)$, the $r-t$ component of  the covariant energy momentum 
 tensor (\ref{9})  becomes,
\begin{eqnarray}
T^{r}_t&=& \frac{e^2}{4\pi}(A_{t}(r) - A_{t}(r_{+}))^2 
+\frac{1}{12\pi}(p - \frac{1}{4}(f' + z))^2 \nonumber\\
&&+\frac{1}{24\pi}(pf' + \frac{1}{4}ff'' - \frac{1}{4}f'(f'+z)).\label{18}
\end{eqnarray}
Now we implement the boundary condition; namely the vanishing of the covariant energy momentum tensor
at the horizon, $T^{r}_t(r_{+}) = 0$. This 
condition yields,
%\begin{equation}
% (p - \frac{1}{4}(f'_{+}+z))^2 + \frac{1}{2} (pf'_{+} - \frac{1}{4}f'_{+}(f'_{+} +z))= 0,
%\end{equation}
\begin{equation}
p=\frac{1}{4}(z \pm f'_{+}) \ ; \  f'_{+} \equiv f'(r = r_{+}). \label{19} 
\end{equation} 
Using either of the above solutions in (\ref{18}) we get,
\begin{eqnarray}
T^{r}_t = \frac{e^2}{4\pi} \left(A_{t}(r) - A_{t}(r_{+})\right)^2\nonumber\\
 + \frac{1}{192\pi} \left[f'^{2}_{+} - f'^{2} + 2ff''\right]. \label{24}
\end{eqnarray} 
This expression is in agreement with that given in \cite{shailesh}. 

To obtain the energy flux, we recall that it is given by the asymptotic expression for the anomaly free energy momentum
tensor. As happened for the charge case, here also it is found from (\ref{13}) that the anomaly vanishes in this limit.
Hence the energy flux is abstracted by taking the asymptotic infinity limit of (\ref{24}). This yields,
\begin{equation}
T^{r}_t(r\rightarrow\infty)=\frac{e^2}{4\pi} A^{2}_{t}(r_{+})+ \frac{1}{192\pi}f'^{2}_{+}.\label{25}
\end{equation}
which correctly reproduces the Hawking flux. 

{\it {Connection with Unruh vacuum}}\\

Here we compute the anomaly free current and the energy momentum tensor, which describe the theory away from the 
horizon, and show that these agree with the expectation values of these observables for the Unruh vacuum.

  We consider the expression for the current $J^{\mu}_{(o)}$ in the region
 outside the horizon. From (\ref{4}) and (\ref{7})  we obtain,
\begin{equation}
J^{r}_{(o)} = \frac{e^2}{\pi}a, \  J^{t}_{(o)} = \frac{e^2}{\pi f}\left(A_{t}(r) + c\right).\label{16}
\end{equation}
At asymptotic infinity the result for $J^{r}_{(o)}$ must agree with (\ref{chargeflux}). Taken together
with (\ref{coeff}) this implies  
 $a = c = -\frac{A_{t}(r_{+})}{2}$ and hence the currents outside the horizon are given by,
\begin{equation}
J^{r}_{(o)} = -\frac{e^2}{2\pi} A_{t}(r_{+}) \ ; \ J^{t}_{(o)} = \frac{e^2}{\pi f}\left(A_{t}(r) - \frac{1}{2}A_{t}(r_{+})\right).\label{17}
\end{equation}    
This is also the expectation value of the current for the Unruh vacuum in the $d=2$ RN black hole \cite{Isoumtwilczek, unruh}.

Now we consider components of the anomaly free energy momentum tensor defined 
 in (\ref{3}). The $r-t$ component of $T^{\mu}_{\nu(o)}$ is given by
\begin{eqnarray}
T^{r}_{t(o)} &=&  \frac{e^2}{4\pi} A^{2}_{t}(r_{+}) - \frac{e^2}{2\pi} A_{t}(r)A_{t}(r_{+})\nonumber\\
&& -\frac{1}{12\pi}zp,\label{20}
\end{eqnarray}
while the $t-t$ component becomes,
\begin{eqnarray}
T^{t}_{t(o)} = \frac{e^2}{2\pi f}\left(A_{t}^2(r) - A_{t}(r_{+})A_{t}(r) + 
 \frac{1}{2}A^{2}_{t}(r_{+})\right) \nonumber\\
+ \frac{1}{48\pi f} \left[2ff''-f'(f'+z) + 8p^2 + \frac{(f'+z)^2}{2}\right].\label{21}
\end{eqnarray}

The asymptotic form of (\ref{20}) must agree with that of (\ref{25}). A simple inspection shows that $zp= -\frac{1}{16}f'^{2}_{+}$.
Substituting this in (\ref{19}) yields two solutions $p = \frac{1}{8}f'_{+}; \ z = -\frac{1}{2}f'_{+}$ 
and $p = -\frac{1}{8}f'_{+}; \  z = \frac{1}{2}f'_{+}$. Using either of these  solutions in (\ref{20}) and 
(\ref{21}) we obtain,
\begin{eqnarray}
T^{r}_{t(o)} &=&  \frac{e^2}{4\pi} A^{2}_{t}(r_{+}) - \frac{e^2}{2\pi} A_{t}(r)A_{t}(r_{+})\nonumber\\
&& +\frac{1}{192\pi}f'^2_+,\label{20a}
\end{eqnarray}
while the $t-t$ component becomes,
\begin{eqnarray}
T^{t}_{t(o)} = \frac{e^2}{2\pi f}\left(A_{t}^2(r) - A_{t}(r_{+})A_{t}(r) + 
 \frac{1}{2}A^{2}_{t}(r_{+})\right) \nonumber\\
+ \frac{1}{96\pi f} \left[4ff''-f'^2 + \frac{1}{2}f'^2_+ \right].\label{21a}
\end{eqnarray}
Likewise  $T^{r}_{r(o)}$can be computed either directly or from noting the trace $T^\mu_{\mu(o)} = \frac{R}{24\pi}$
that follows from(\ref{3}) and then using (\ref{21a}).
These are also the expressions for the expectation values of the various components of the energy momentum tensor
found for the Unruh vacuum \cite{ unruh,Isoumtwilczek}.

{\it Discussions:}\\

 We have given a derivation of the Hawking flux from charged black holes, based on the effective action approach,
which only employs the boundary conditions at the event horizon. It might be mentioned that generally such approaches
require, apart from conditions at the horizon, some other boundary condition, as for example, the vanishing of ingoing modes at 
infinity \cite{Fulling, wipf, unruh}. The latter obviously goes against the universality of the Hawking effect which should 
be determined from conditions at the horizon only. In this we have succeeded. Also, the specific structure of the effective action
from which the Hawking radiation is computed is valid only at the event horizon. This is the anomalous (chiral)
effective action. Other
effective action based techniques do not categorically specify the structure of the effective action at the horizon.
Rather, they use the usual (anomaly free) form for the
effective action and are
restricted to two dimensions only \cite{wipf}.

  An important factor concerning this analysis is to realise that effective
field theories become two dimensional and chiral near the event horizon
\cite{Robwilczek}. Yet another ingredient was the implementation
of a specific boundary condition- the vanishing of the covariant form of the current and/or the energy
momentum tensor \cite{Isowilczek, shailesh}. Not only that, the importance of the covariant forms were
further emphasised by us in \cite{shailesh} where it was shown that the anomaly cancelling approach was simplified
considerably if, instead of consistent anomalies used in \cite{Robwilczek, Isowilczek}, covariant anomalies
were taken as the starting point. Indeed, in the present computations, we have taken that form of the effective action which yields anomalous Ward identities 
having covariant gauge and gravitational anomalies. 
The unknown parameters in the covariant energy momentum tensor and the covariant current 
derived from this anomalous effective action were fixed by a boundary condition- namely the vanishing of these 
covariant quantities
at the event horizon. Consequently we have shown that aspects like covariant anomalies and covariant boundary
conditions 
are not merely confined to
discussing the Hawking effect in the anomaly cancelling approach \cite{Robwilczek, Isowilczek, shailesh}.
Rather they have a wider applicability since our effective action based approach is different from the anomaly
cancelling approach. 

Further, we have exploited the information from  the chiral (anomalous) 
effective action, which describes the theory near the horizon,
to completely fix the form of the normal effective action that describes the theory away from the horizon. The 
expressions for the currents and energy momentum tensors obtained from the latter reproduce the results obtained
by using the Unruh vacuum approach \cite{unruh, Isoumtwilczek}. There is an alternative approach, discussed
in the appendix of  \cite{Isoumtwilczek},
that reveals the connection of the normal effective action with Unruh vacuum. However it uses the 
Kruskal coordinates and directly imposes, as a boundary condition,  the vanishing
of ingoing modes at infinity. Hence it is different from our approach.

{\it Acknowledgements:}\\
The authors thank Saurav Samanta for discussions. 
One of the authors (RB) also thanks members of the Institute of
 Quantum Science, Nihon University, Tokyo, where a part
of this work was done, for their gracious hospitality and support, and 
Satoshi Iso for fruitful discussions.     
%%%%%%%%%%%%%%%%%%%%%%%%%%%%%%%%%%%%%%%%%%%%%%%%%%%%%%%%%%%%%%%%%%%%%%%%%%

%%%%%%%%%%%%%%%%%%%%%%%%%%%%%%%%%%%%%%%%%%%%%%%%%%%%%%%%%%%%%%%%%%%%%%%%%%
\end{document}